\documentclass[conference,letterpaper,10pt]{IEEEtran}
\usepackage{cite}
\usepackage{amsmath,amssymb,amsfonts}
\usepackage{graphicx}
\usepackage{textcomp}
\usepackage{booktabs}
\usepackage{array}
\usepackage{tabularx}
\usepackage{url}
\usepackage{microtype}
\usepackage{tikz}
\usepackage{paralist}
\usepackage{xcolor}

\def\BibTeX{{\rm B\kern-.05em{\sc i\kern-.025em b}\kern-.08em
    T\kern-.1667em\lower.7ex\hbox{E}\kern-.125emX}}
\newcolumntype{P}[1]{>{\raggedright\arraybackslash}p{#1}}

\newcommand\copyrighttext{%
\footnotesize This is the preprint accepted for publication in the 4th International Conference on Federated Learning Technologies and Applications (FLTA 2026), Paris, France, 27-30 October 2026. The final version will be available in the proceedings of the conference that will be published by IEEE.}
\newcommand\ieeecopyrighttext{%
\scriptsize \textcopyright\ 2026 IEEE. Personal use of this material is permitted. Permission from IEEE must be obtained for all other uses, in any current or future media, including reprinting/republishing this material for advertising or promotional purposes, creating new collective works, for resale or redistribution to servers or lists, or reuse of any copyrighted component of this work in other works.}
\newcommand\copyrightnotice{%
\begin{tikzpicture}[remember picture,overlay]
\node[anchor=north,yshift=-10pt] at (current page.north) {\fbox{\parbox{\dimexpr\textwidth-\fboxsep-\fboxrule\relax}{\copyrighttext}}};
\node[anchor=south,yshift=10pt] at (current page.south) {\parbox{\textwidth}{\ieeecopyrighttext}};
\end{tikzpicture}%
}

\begin{document}

\title{AuditFL: A Contract-Backed Clearing Layer for Open Federated Learning}

\author{
\IEEEauthorblockN{Alexandros Balla}
\IEEEauthorblockA{
Department of Computer Science and Engineering\\
Hong Kong University of Science and Technology\\
Hong Kong SAR, China\\
Email: aballa@cse.ust.hk
}
\and
\IEEEauthorblockN{Spyros Lalis}
\IEEEauthorblockA{
Department of Electrical and Computer Engineering\\
University of Thessaly\\
Volos, Greece\\
Email: lalis@uth.gr
}
}

\maketitle

\copyrightnotice

\begin{abstract}
Federated learning (FL) keeps raw data local, but most deployments still rely on a trusted coordinator for client admission, checkpoint distribution, update collection, aggregation, and rewards. This paper presents AuditFL, a contract-backed clearing layer for auditable FL in open settings. AuditFL keeps training and aggregation off chain, stores model artifacts in an assumed reliable off-chain artifact store, and records only content identifiers and economic state on chain. Clients reserve update tickets with deposits, publish update identifiers, and later claim refunds or rewards; the task coordinator publishes global checkpoints and finalizes rounds through deterministic settlement rules. The design does not remove the aggregator, but makes its on-chain round decisions attributable, replayable, auditable, and economically constrained. We implement AuditFL as both a deterministic state-machine prototype and a Solidity contract on a local Ethereum-compatible Anvil chain. On full MNIST with Dirichlet non-IID splits, 12 rounds, and up to 200 clients, AuditFL preserves the Plain FL learning trajectory while its overhead follows closed-form control-plane models. At 200 clients, the contract run issues 5,026 transactions, stores 2,413 off-chain artifacts (144.9 MiB), and consumes 829.7M EVM gas, dominated by per-client reserve and publish calls. Controlled dropout, invalid-update, and artifact-unavailability stress tests reduce inclusion nearly linearly but still finalize rounds, with final-accuracy changes remaining within a few tenths of an accuracy percentage point even under high injected failure rates. 
\end{abstract}

\begin{IEEEkeywords}
Federated learning, blockchain, smart contracts, permissionless systems, auditable coordination. 
\end{IEEEkeywords}

\section{Introduction}
Federated learning (FL) trains a shared model by sending computation to clients instead of moving raw data to a central server~\cite{McMahan2016CommunicationEfficientLO,kairouz2021advancesopenproblemsfederated,bonawitz2019towards,li2020federatedchallenges}. This is useful for edge devices, healthcare, finance, and cross-organization learning, where data movement is expensive, sensitive, or restricted. However, the standard FL control plane is usually centralized. A coordinator selects clients, distributes checkpoints, receives updates, aggregates them, records progress, and often controls rewards. If this coordinator is compromised or biased, it can censor participants, hide updates, manipulate aggregation, or publish a round outcome whose inputs, checkpoint, and accounting cannot be independently reconstructed.

Blockchains offer a public ordering layer for distributed protocols through replicated logs, fork choice, and consensus assumptions~\cite{nakamoto2008bitcoin,garay2015backbone}.  
In FL, a ledger can be used to record task creation, update commitments, round transitions, deposits, and rewards. A naive blockchain design is not sufficient, though. If every model tensor is stored or executed on chain, the ledger becomes a slow storage and computation 
bottleneck. If the ledger is permissioned, the system cannot support an open market where any client may join. If the consensus protocol is left unspecified, Sybil resistance and finality assumptions are unclear; this is especially important because open peer-to-peer systems are classically vulnerable to identity multiplication attacks~\cite{douceur2002sybil}. 

These gaps are visible across blockchain-assisted FL surveys and systems, which often emphasize incentives, privacy, or application frameworks while leaving consensus assumptions and round-level state-machine details less explicit~\cite{kim2019blockchained,lu2020blockchain,shayan2021biscotti,qu2022blockchainenabled,huang2023systematicbfl,wang2024systematicsurveyblockchainedfederated}.

This paper presents AuditFL, a contract-backed clearing layer for open FL. The ledger orders and validates small state transitions, while model bytes remain in an assumed reliable distributed off-chain storage layer and are referenced by content identifiers. A task owner, called the \emph{task coordinator}, applies the configured aggregation rule and publishes the next global checkpoint. AuditFL keeps aggregation with this designated actor, but moves round ordering, commitments, deposits, refunds, rewards, and claims into a public clearing layer. The resulting trace makes coordinator actions visible, replayable, auditable, and economically constrained.

The paper makes four contributions:
\begin{inparaenum}[(i)]
\item We define AuditFL as a contract-backed clearing layer for open FL, separating on-chain coordination and settlement from off-chain training, aggregation, and artifact storage.
\item We specify the round-clearing state machine, including task listing, funding, ticket reservation with deposits, update publication, global finalization, refunds, reward entitlements, and claims.
\item We give a layered safety and liveness argument showing how ledger ordering, confirmation depth, ticket expiry, deposits, refunds, and scoring make coordinator and client actions replayable and economically accountable under partial synchrony~\cite{dwork1988partiallysynchronous}.
\item We implement and evaluate the clearing layer on a local Ethereum-compatible chain with a non-IID MNIST workload, measuring transactions, gas cost, artifact counts and bytes, and behavior under dropout, invalid-update, and artifact-unavailability stress.
\end{inparaenum}

\section{Related Work and Positioning}

\begin{table*}[t]
\centering
\caption{Positioning against representative blockchain-assisted, accountable, auditable, and verifiable FL work.}
\label{tab:relatedcomparison}
\scriptsize
\setlength{\tabcolsep}{2.2pt}
\begin{tabular}{l c c c c c c l}
\toprule 
\textbf{Work} & \textbf{Open} & \textbf{Commit.} & \textbf{Partic.} & \textbf{Econ.} & \textbf{Contract} & \textbf{Fault/} & \textbf{Main emphasis} \\
 & \textbf{admit.} & \textbf{refs.*} & \textbf{bond} & \textbf{settle.} & \textbf{cost} & \textbf{attack focus} & \\
\midrule

BlockFL~\cite{kim2019blockchained}
& $\triangle$ & -- & -- & -- & -- & -- & block-generation latency \\

Biscotti~\cite{shayan2021biscotti}
& $\checkmark$ & $\checkmark$ & -- & -- & -- & $\checkmark$ & private/secure P2P FL \\

FedCoin~\cite{liu2020fedcoin}
& $\triangle$ & -- & -- & $\checkmark$ & -- & -- & Shapley-based payments \\

ChainFL~\cite{korkmaz2020chainfl} 
& -- & -- & -- & -- & $\triangle$ & -- & on-chain model storage/aggregation \\

BlockFLA~\cite{desai2021blockfla}
& -- & $\checkmark$ & $\checkmark$ & $\checkmark$ & $\checkmark$ & $\checkmark$ & backdoor accountability \\

BlockFLow~\cite{mugunthan2020blockflow}
& -- & $\checkmark$ & $\checkmark$ & $\checkmark$ & $\checkmark$ & $\checkmark$ & accountable rewards \\

Auditable BR-FL~\cite{auditableByzantineFL} 
& -- & $\checkmark$ & $\triangle$ & $\triangle$ & -- & $\checkmark$ & Byzantine-robust audit \\

VerifBFL~\cite{bellachia2025verifbfl}
& -- & $\checkmark$ & $\checkmark$ & $\checkmark$ & $\triangle$ & $\checkmark$ & zk-verifiable FL \\

\textbf{AuditFL}
& $\checkmark$ & $\checkmark$ & $\checkmark$ & $\checkmark$ & $\checkmark$ & $\checkmark$ & auditable round clearing \\

\midrule

\multicolumn{8}{c}{$\checkmark$ = explicit focus; $\triangle$ = partial or adjacent support; -- = not central in the cited work.}\\

\multicolumn{8}{c}{* Cryptographic commitments or content identifiers for model/update objects; open admission excludes pre-approved clients.}\\

\bottomrule
\end{tabular}
\end{table*}

FL was introduced as a practical way to train on decentralized client data using iterative model averaging~\cite{McMahan2016CommunicationEfficientLO}. Later system work clarified that real FL deployments require scheduling, retry logic, guided participant selection, benchmarking infrastructure, secure aggregation, device availability handling, production control-plane engineering, and robustness to statistical and systems heterogeneity~\cite{bonawitz2019towards,kairouz2021advancesopenproblemsfederated,li2020federatedchallenges,li2020fedprox,lai2021oort,lai2022fedscale}. Those works do not attempt to make client admission or round finalization permissionless. They usually assume an operator-controlled service. Our work keeps the FL computation model, but moves task creation, update admission, round transitions, and accounting into a public ledger state machine.

BlockFL studies the latency consequences of replacing a central FL coordinator with block generation~\cite{kim2019blockchained}. Biscotti uses blockchain and cryptographic primitives for private and secure peer-to-peer FL, while FedCoin and ChainFL explore blockchain-based incentives or decentralized model coordination~\cite{shayan2021biscotti,liu2020fedcoin,korkmaz2020chainfl}. BlockFLA is closest to our accountability motivation: it records SHA-256 hashes, deposits, and penalties for detected backdoor attackers through a hybrid blockchain design~\cite{desai2021blockfla}. 

Accountable, auditable, and verifiable FL systems address adjacent trust problems. BlockFLow uses smart contracts and auditing mechanisms to reward agents based on contribution quality~\cite{mugunthan2020blockflow}. Auditable Byzantine-robust FL studies auditability at the robust-aggregation layer~\cite{auditableByzantineFL}, while verifiable-FL systems target cryptographic checks of aggregation or training correctness~\cite{fu2020vfl,bellachia2025verifbfl}. AuditFL is complementary: it focuses on the round-clearing control plane, namely ticketed admission, update commitments, deterministic settlement of the submitted finalization record, refund/claim settlement, and measured contract execution cost.

Other BC-FL systems focus on privacy-preserved data sharing in Industrial IoT or secure/incentive FL settings~\cite{lu2020blockchain,Xu21bcfl}. Surveys confirm that blockchain-assisted FL has been studied across decentralization, incentives, storage, privacy, security, and application settings, while scalability, storage pressure, consensus assumptions, and adversarial participation remain recurring open issues~\cite{qu2022blockchainenabled,huang2023systematicbfl,wang2024systematicsurveyblockchainedfederated,ning2024bflsurveynewperspectives,zhang2024decentralizedbfl,hallaji2024dflsecurity}.

Table~\ref{tab:relatedcomparison} positions AuditFL against representative systems. The comparison is feature-oriented: it records whether a work explicitly emphasizes a feature as part of its design or evaluation, not whether that feature could be added. Unlike accountable, Byzantine-robust, or verifiable FL systems that audit contribution quality or aggregation correctness, AuditFL audits the clearing layer around each round: ticketed admission, update commitments, deterministic settlement of the submitted finalization record, refunds, claims, and contract execution cost.

\section{System Model and Design Goals}

\subsection{Roles and Assumptions}
The system has three active roles. The \emph{task coordinator} creates the FL task, funds the reward pool, retrieves revealed updates, applies the configured aggregation rule, publishes the new global checkpoint, and finalizes each round. This is the conventional FL aggregator role; AuditFL does not remove it, but makes its round-level actions public, replayable, and economically constrained. \emph{Clients} read the current model state, reserve tickets, train on local data, make update artifacts available through the assumed off-chain artifact store, publish update content identifiers on chain, and claim any resulting refund or reward. The \emph{ledger layer} provides authenticated ordering and deterministic validation for signatures, balances, nonces, round indices, ticket ownership, deadlines, rate limits, identifier syntax, and state transitions.

The task coordinator is designated but auditable: AuditFL decentralizes participation, ordering, and accounting, while keeping aggregation execution with a visible task owner whose round-level actions are recorded in the public trace.

We assume partial synchrony for a deployed permissionless ledger: before an unknown stabilization time, messages may be delayed, but after stabilization honest messages are delivered within a bounded delay~\cite{dwork1988partiallysynchronous}. The adversary may create many client identities, submit malformed content identifiers, withhold off-chain artifacts, attempt model poisoning, misreport metrics, and, in a proof-of-stake (PoS) deployment, corrupt less than half of the bonded stake. A task coordinator may also behave strategically by omitting available updates or using a weak scoring rule. The adversary cannot forge standard digital signatures or find collisions in the content hashes used by identifiers. Public consensus security is a deployment assumption; AuditFL specifies the clearing-layer rules that execute above a ledger satisfying these assumptions.

Threats are separated by layer. Consensus faults may delay, censor, fork, or reorder ledger events within the assumed consensus bounds. Clearing-layer faults include malformed identifiers, low-quality updates, unavailable artifacts, selective omission, or weak scoring. These faults affect inclusion, settlement, and auditability, but not block validity: invalid or unavailable updates are skipped and settled by policy, and the ledger never fetches tensors.

\subsection{Design Goals and Non-Goals}
The first goal is \emph{open coordination}: clients and validators should not need prior admission by a central authority. The second goal is \emph{minimal consensus}: block validation should not depend on ML libraries, GPU kernels, or off-chain storage fetches. The third goal is \emph{round auditability}: from public chain state and retrievable content identifiers, an auditor should be able to replay the round-level control flow and verify the economic outcome. The fourth goal is \emph{economic spam resistance}: wasting publication slots or withholding content should cost deposits or fees, aligning with recent incentive-oriented FL and BC-FL discussions~\cite{zhan2021incentivemechanism}.

AuditFL is not a complete privacy or poisoning-defense mechanism. Raw data remain local, but model updates may still leak information, and malicious updates may still affect the learned model. Stronger guarantees require complementary mechanisms such as secure aggregation~\cite{bonawitz2017practical}, differential privacy~\cite{kairouz2021advancesopenproblemsfederated}, and robust aggregation~\cite{blanchard2017,yin2021byzantine,Pillutla_22_robust_aggregation}.

\begin{figure*}[!t]
\centering
\includegraphics[width=0.84\textwidth]{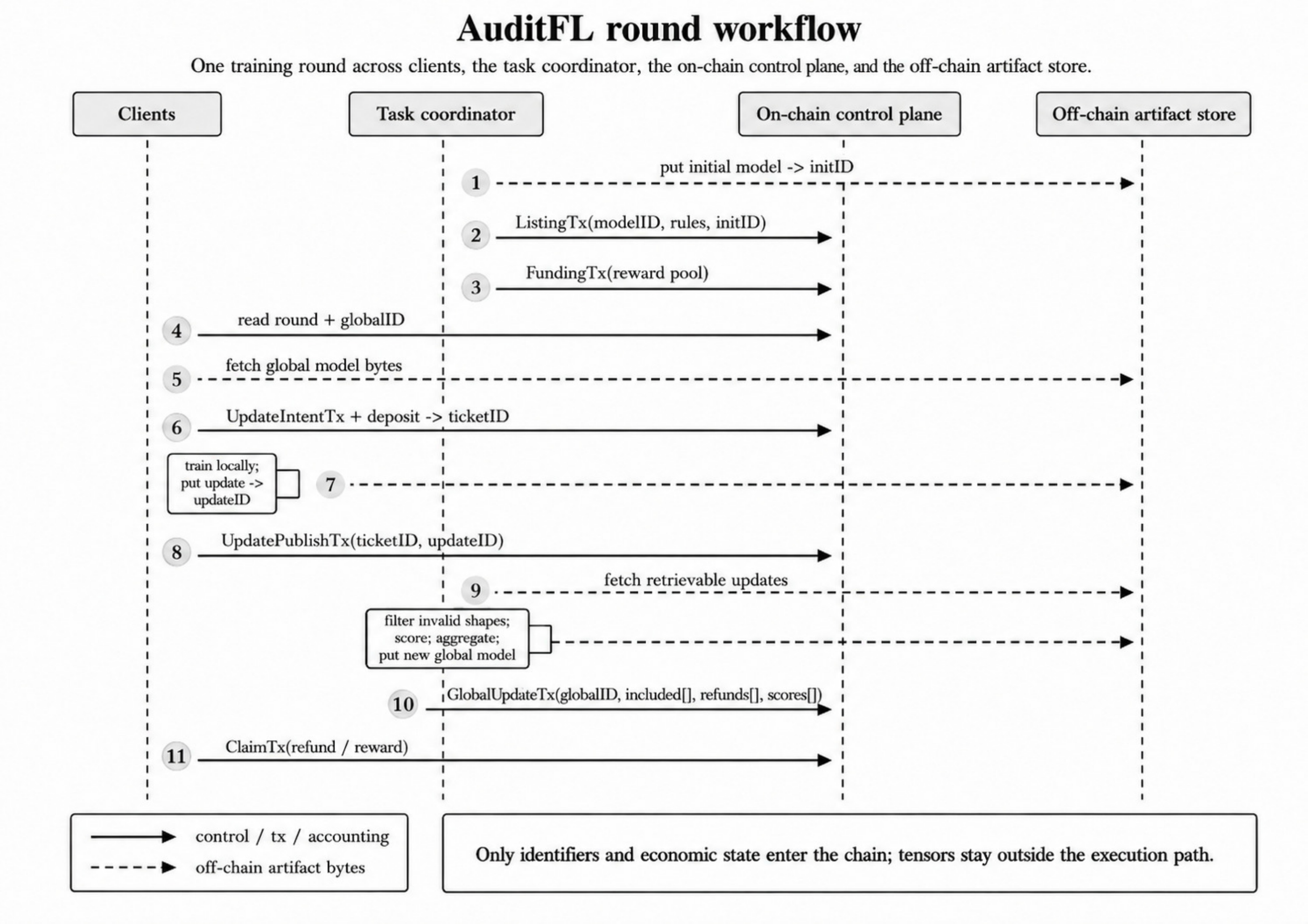}
\caption{AuditFL round workflow. Solid arrows denote control transactions and accounting state, while dashed arrows denote off-chain artifact transfers. Only identifiers and economic state enter the chain; tensors stay outside contract execution.}
\label{fig:architecture}
\end{figure*}

\section{Protocol}

\subsection{Round Workflow}
Fig.~\ref{fig:architecture} summarizes one training round. The task coordinator first uploads an initial checkpoint to off-chain storage and submits a \texttt{ListingTx} that binds a \texttt{modelID} to an initial content identifier and an immutable rule set, including deadlines, deposits, and the aggregation/scoring policy. The task coordinator then funds the reward pool with \texttt{FundingTx}. A client reads the current round, locks a deposit through \texttt{UpdateIntentTx}, receives a ticket, trains locally, uploads the update to off-chain storage, and binds the ticket to an update identifier through \texttt{UpdatePublishTx}. At the deadline, the task coordinator fetches retrievable revealed updates, filters invalid shapes, scores candidates, aggregates eligible updates, uploads the new global checkpoint, and submits \texttt{GlobalUpdateTx}. The ledger state machine then advances the round, records the latest global identifier, credits rewards, and marks refundable deposits, which clients withdraw through \texttt{ClaimTx}.

The key separation is between ordering, artifact retrieval, and learning. The ledger orders and validates round-level state transitions; the off-chain artifact layer serves model bytes; and training, scoring, validation, and aggregation execute off chain according to the task rules. Invalid or unavailable updates are skipped and settled by policy rather than blocking round progress.

\subsection{Transaction State Machine}
Table~\ref{tab:tx} lists the transaction objects. All validators execute the same deterministic state machine. A model must be listed before it can be funded or updated. A publication must reference a live unrevealed ticket created by an earlier intent for the same model and round. A global update must match the current round and can only reference tickets that are already revealed. These invariants prevent hidden updates, double publication, stale finalization, and unreserved update spam.

Validators also apply a strict replay rule. Admission checks are useful for mempool hygiene, but block validity is decided by re-executing the ordered transactions from the parent state. A block that admits a transaction in the mempool can still be invalid if another transaction in the same block consumes the relevant balance, advances the round, or reveals the ticket first. This is why the protocol treats every state transition as atomic and why repeated calls use nonces and idempotent checks.

\begin{table}[t]
\centering
\caption{AuditFL transaction objects.}
\label{tab:tx}
\scriptsize
\setlength{\tabcolsep}{2.5pt}
\begin{tabular}{p{0.25\columnwidth}p{0.67\columnwidth}}
\toprule
\textbf{Object} & \textbf{Main effect and checks} \\
\midrule
\texttt{ListingTx} & Creates a task with model identifier, initial content identifier, deadlines, deposits, and immutable rules. \\
\texttt{FundingTx} & Adds tokens to the reward pool after signer, balance, and amount checks. \\
\texttt{UpdateIntentTx} & Reserves a ticket, locks a deposit, and enforces per-account caps. \\
\texttt{UpdatePublishTx} & Binds a live unrevealed ticket to an update content identifier. \\
\texttt{GlobalUpdateTx} & Finalizes the round with global identifier, included tickets, refunds, and scores; advances the round and credits entitlements. \\
\texttt{ClaimTx} & Debits an entitlement and credits the claimant balance. \\
\bottomrule
\end{tabular}
\end{table}

\begin{figure}[t]
\small
\setlength{\fboxsep}{4pt}
\fbox{\begin{minipage}{0.94\columnwidth}
\textbf{Round finalization rule.} For a fixed task and round \(r\): 
(1) read all round-\(r\) tickets from the \(k\)-deep chain view and identify the revealed subset; 
(2) sort revealed tickets by identifier; 
(3) fetch referenced updates with bounded retries; 
(4) add usable revealed tickets to \texttt{refunds[]}; 
(5) add shape-valid and score-positive updates to \texttt{included[]} and \texttt{scores[]}; 
(6) aggregate included updates off chain; 
(7) upload the global checkpoint and submit \texttt{GlobalUpdateTx}; 
(8) the ledger advances \(r\) only if referenced tickets are valid, submitted arrays align, and unresolved tickets are settled according to policy.
\end{minipage}}
\caption{Round finalization rule. The task coordinator aggregates off chain; the ledger checks identifiers, array alignment, and settlement state, not tensor bytes.}
\label{fig:finalize}
\end{figure}

\subsection{Tickets, Refunds, and Payouts}
\label{subsec:payouts}

A ticket reserves one update-publication slot for the current round and requires the client to lock a deposit. If a client reserves a ticket but never publishes an update identifier, the deposit is forfeited. If the client publishes an identifier whose artifact is retrievable and passes the rule-declared serialization and shape checks, the deposit is refundable even when the update earns no reward. If the artifact is unavailable or fails these basic validity checks, the deposit is forfeited. Thus, the deposit prices reservation and unusable-update behavior, while reward payments depend on the contribution score.

The task coordinator assigns a nonnegative score \(s_j\) to each included update \(j\). One illustrative rule is
\begin{equation}
  s_j = \max(0,\Delta_j) w_j,
\end{equation}
where \(\Delta_j\) is the improvement of candidate update \(j\) over the previous global model on the task coordinator's validation set, and \(w_j\) is a capped proxy for the client's local sample size. For a fixed task, let \(R_r\) be the reward pool for round \(r\), let \(I_r\) be the set of updates included in that round, and let \(\gamma \in [1,2]\) be the listing-time payout exponent, with larger values concentrating rewards on higher-scoring updates. The payout for an included update \(j \in I_r\) is
\begin{equation}
  \mathrm{pay}_j =
  R_r\frac{s_j^{\gamma}}{\sum_{u\in I_r}s_u^{\gamma}},
\end{equation}
with zero payout when \(s_j=0\) or the denominator is zero.

The ledger enforces settlement arithmetic rather than the correctness of the task-specific scoring policy. It applies the submitted score vector \(\mathbf{s}=(s_j)_{j\in I_r}\), whose entries correspond to the included ticket identifiers in the finalized round, together with the declared rule parameters. When the listing exposes a public validation set or a committed evaluation procedure, an auditor can recompute scores and reward entitlements; otherwise, the auditor can verify only that payouts, refunds, forfeitures, and claims are consistent with the submitted scores and on-chain rules. Within a confirmed round, every funded reward and locked deposit is accounted for through a reward entitlement, refund entitlement, forfeiture destination, remaining pool balance, or explicitly charged fee.

\section{Ledger and Artifact Assumptions}

\subsection{PoS Longest-Chain Deployment Model}
For a permissionless proof-of-stake (PoS) ledger, time is divided into slots. In slot \(t\), validator \(i\) with stake \(\mathrm{st}_i\) and total stake \(\mathrm{ST}\) may be eligible to propose a block according to stake-weighted randomness:
\begin{equation}
  \Pr[i\text{ proposes a block in slot }t]\approx \tau\frac{\mathrm{st}_i}{\mathrm{ST}},
\end{equation}
where \(\tau\) controls the expected proposer density. Splitting the same stake across many identities does not increase expected eligibility, because leader probability follows total stake rather than identity count. If multiple valid blocks are observed, fork choice selects the longest valid chain, with a deterministic tie-breaker.

Protocol participants that depend on a ledger write, such as clients waiting for ticket confirmation or the task coordinator waiting to finalize a round, treat the write as stable after \(k\) confirmations. Larger \(k\) reduces reorganization risk but increases round latency. If a round depends on \(B_r\) block-producing stages, the expected latency can be approximated as
\begin{equation}
  L_{\mathrm{round}} \approx (B_r+k)\frac{\Delta}{p(n,\tau)},
\end{equation}
where \(\Delta\) is slot duration, \(n\) is the number of equal-stake validators, and
\(p(n,\tau)=1-(1-\tau/n)^n\) is the probability that at least one validator is eligible in a slot. This model is intentionally simple, but it explains why lower confirmation depth and moderate \(\tau\) reduce latency, while high proposer density increases fork contention.

AuditFL only requires the ledger interface exposed by this model: open transaction submission, stake-weighted Sybil resistance, an ordered log of confirmed state transitions, and probabilistic finality. The clearing-layer protocol does not depend on a specific PoS implementation; a public longest-chain ledger, a chain with BFT finality, or another ledger providing authenticated ordering and stable confirmed writes can instantiate the same transaction model.

\subsection{Off-Chain Storage Interface}

AuditFL assumes a reliable distributed off-chain storage layer that stores and retrieves model artifacts by content identifier during the relevant round window. The ledger never stores tensors and never fetches artifact bytes during validation. It only checks that submitted identifiers are syntactically well formed and bound to the correct ticket or round transition. Clients, task coordinators, and auditors retrieve artifacts outside consensus and verify that the bytes match the published identifiers. Systems such as IPFS~\cite{benet2014ipfs} illustrate one possible substrate, but the storage protocol itself is outside the scope of AuditFL.

This interface has two advantages. First, blocks stay small because they contain content identifiers, ticket IDs, scores, and state deltas rather than tensors. Second, slow or missing off-chain content does not halt consensus. If an update cannot be retrieved before aggregation closes, the task coordinator omits it and the refund policy decides the economic outcome.

\section{Security and Correctness Discussion}

The consensus layer targets common prefix, chain growth, and chain quality under partial synchrony and honest majority of stake~\cite{garay2015backbone,kiayias2017ouroboros}. Confirmation depth prevents the FL layer from acting on tentative history. If a publication or global update is displaced before depth \(k\), the caller retries against the new canonical state. Transaction nonces and idempotent checks prevent duplicated state effects.

Clearing-layer safety follows from causality invariants. A valid \texttt{GlobalUpdateTx} cannot name an unlisted model, cannot finalize an old round, and cannot include tickets that were not revealed. Economic safety follows from explicit pools and entitlements: deposits and rewards move only through deterministic transfers, and the settlement invariant in Sec.~\ref{subsec:payouts} accounts for each funded reward and locked deposit. Validator equivocation can be made slashable because conflicting signed messages provide evidence. Transaction fees, locked deposits, and intent caps make reservation spam costly and bound outstanding ticket reservations; they do not, by themselves, guarantee Sybil-proof reward fairness or truthful contribution scoring.

Learning integrity is intentionally separated from consensus. Byzantine clients can still submit poisoned tensors, and the task coordinator can choose weak validation rules or selectively include updates. AuditFL can enforce basic well-formedness checks, such as serialization and shape checks, while leaving robust aggregation to the task-specific aggregation rule. Stronger aggregation rules such as Krum, coordinate-wise median, trimmed mean, or recent robust FL methods can be inserted without changing the ledger state machine~\cite{blanchard2017,yin2021byzantine,Pillutla_22_robust_aggregation}. Privacy is also model-dependent. Raw data never enter the ledger, but updates can leak information through gradients or membership signals; secure aggregation and differential privacy remain compatible model-specific options~\cite{zhu2019deep,melis2019featureleakage,shokri2017membership,bonawitz2017practical,kairouz2021advancesopenproblemsfederated}.

A complete round has a replay path. An auditor starts from the confirmed listing rules, previous global content identifier, and revealed tickets. For each included update, the auditor fetches the referenced artifact, checks serialization and shape rules, recomputes public scores when the validation data are available, reruns the deterministic aggregator, and checks that the resulting bytes hash to the accepted \texttt{globalCID} or equivalent identifier. The auditor can also recompute refund and reward entitlements from \texttt{refunds[]} and \texttt{scores[]}. This is weaker than a cryptographic proof of correct aggregation, but stronger than trusting a silent coordinator: deviations leave public evidence.

AuditFL narrows, rather than removes, trust in the task coordinator. It does not prove that the selected aggregation outcome is optimal, but it prevents silent history rewrites, hidden updates, and arbitrary reward minting because round transitions and payouts are anchored in the ledger. When validation data or a committed evaluation oracle are available, scoring can also be replayed. Thus, the security claim is detectability and accountability, not cryptographic enforcement of aggregation correctness.

\section{Prototype Implementation}

In our prototype\footnote{An artifact repository containing the Solidity contract, experiment runners, plotting scripts, and paper-result CSVs is available at 

\url{https://anonymous.4open.science/r/AuditFL-6BE3}.}, we instantiate the ledger interface as a Solidity smart contract executed on a local Ethereum-compatible Anvil chain. This evaluates the application-level clearing logic and contract execution cost, not public-chain consensus behavior.

AuditFL also includes a deterministic in-memory state-machine backend used for fast debugging and controlled tests. The Solidity backend deploys a \texttt{BCFLCoordinator} contract that implements the core coordination calls: \texttt{createListing}, \texttt{fundRound}, \texttt{reserveTicket}, \texttt{publishUpdate}, \texttt{finalizeRound}, and \texttt{claim}. The contract stores only coordination metadata, ticket state, reward-pool state, content identifiers, and entitlements.

In Ethereum-compatible chains, smart contracts execute inside the Ethereum Virtual Machine (EVM), while \texttt{gas} is the unit used to meter computational and storage work performed by a transaction~\cite{wood2014ethereum}. We report gas as a proxy for smart-contract execution cost, not as a monetary deployment cost, because monetary cost depends on the gas price of a particular network. 

The client and task coordinator processes are written in Python. Clients read the current round, reserve tickets, train local models, serialize updates, place update bytes in a local SHA-256 content-addressed artifact store~\cite{fips1804}, and publish the resulting 
identifier. The task coordinator fetches retrievable updates, filters invalid shapes, aggregates valid updates using the same FedAvg-style rule as the Plain FL baseline~\cite{McMahan2016CommunicationEfficientLO}, writes the next global model to the artifact store, and finalizes the round on chain. 

\section{Evaluation}

We evaluate AuditFL as a control-plane system along five dimensions:
\begin{inparaenum}[(i)]
\item end-to-end execution;
\item agreement between measured transaction counts and a simple analytical model;
\item EVM execution cost, measured in gas;
\item off-chain artifact-volume scaling; and
\item preservation of the Plain FL learning trajectory.
\end{inparaenum}
We use two backends: a Solidity \texttt{BCFLCoordinator} contract on a local Ethereum-compatible Anvil chain for transaction and gas measurement, and a deterministic state-machine backend for controlled failure stress. Both use the same Python client and task coordinator logic and the same local SHA-256 content-addressed artifact store.

We run experiments on the full 70,000-MNIST dataset loaded through OpenML \cite{lecun1998gradient,vanschoren2014openml}, varying the number of clients
\(N \in \{1,2,5,10,15,20,30,40,50,75,100,150,200\}\).
For each run, we construct a stratified split with 42,000 training examples, 10,500 validation examples, and 17,500 held-out test examples. Client data are drawn only from the training partition; the task coordinator uses the validation partition for update scoring, and final accuracy is reported on the held-out test partition. Each run performs \(R=12\) FL rounds. For \(N>1\), the training partition is distributed non-IID among clients using a Dirichlet distribution with \(\alpha=0.4\). For every client count, Plain FL and AuditFL use the same split, model, local-training rule, number of rounds, and aggregation rule. Thus, any difference between the two runs comes from AuditFL coordination and settlement rather than from the learning rule. The measured on-chain call structure depends on rounds, participating clients, identifiers, and settlement operations, while larger learning workloads primarily affect off-chain computation and, when model size changes, artifact volume.

\subsection{Overhead Model}

Let \(N\) be the number of clients and \(R\) the number of rounds. Plain FL transfers model and update tensors, but does not maintain a public economic log. AuditFL leaves tensors off chain and adds a ledger control plane. In a clean full-participation round, the application transactions are
\begin{equation}
  T_{\mathrm{round}}
  =
  1_{\mathrm{fund}}
  + N_{\mathrm{reserve}}
  + N_{\mathrm{publish}}
  + 1_{\mathrm{finalize}}
  =
  2N+2.
\end{equation}
For a complete run with one listing transaction, \(R\) rounds, and one claim per client, the application transaction count is
\begin{equation}
  T_{\mathrm{app}} = 1 + R \times (2N+2) + N.
\end{equation}
The measured receipt count additionally includes one contract-deployment transaction:
\begin{equation}
  T_{\mathrm{measured}} = 1 + R \times (2N+2) + N + 1.
\end{equation}

Off-chain artifact writes are also predictable. Each clean run stores one initial global model, \(R  \times N\) client-update artifacts, and \(R\) updated global-model artifacts:
\begin{equation}
  P_{\mathrm{put}} = 1 + R \times (N+1).
\end{equation}
This is the artifact-count quantity compared against the clean scaling measurements.

\subsection{Full-MNIST Scaling Results (clean runs, no failures)}

Measured receipt counts match the analytical model for every tested client count. With \(R=12\), the receipt count increases from 51 transactions at \(N=1\) to 5,026 at \(N=200\) clients, matching \(2+R \times(2N+2)+N\). Stored artifact counts also match the analytical model exactly, increasing from 25 artifacts at \(N=1\) to 2,413 at \(N=200\), matching \(1+R\times(N+1)\).

\begin{figure}[!t]
\centering
\includegraphics[width=0.95\columnwidth]{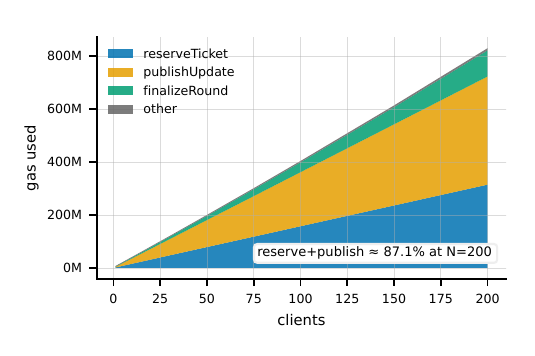}
\caption{Contract execution cost on MNIST with \(R=12\).}
\label{fig:gas_breakdown}
\end{figure}

Fig.~\ref{fig:gas_breakdown} reports the gas breakdown. At \(N=200\), the contract run consumes 829.7M gas: \texttt{publishUpdate} accounts for 408.7M gas and \texttt{reserveTicket} for 314.3M gas, together accounting for approximately 87.1\% of total execution cost. This follows directly from the protocol structure, since each participating client reserves one ticket and publishes one update identifier per round, whereas \texttt{finalizeRound} is invoked once per round. As a further consistency check, the implementation explicitly verifies that the settlement invariant in Sec.~\ref{subsec:payouts} 
holds in every 
run.

\begin{figure}[!t]
\centering
\includegraphics[width=0.95\columnwidth]{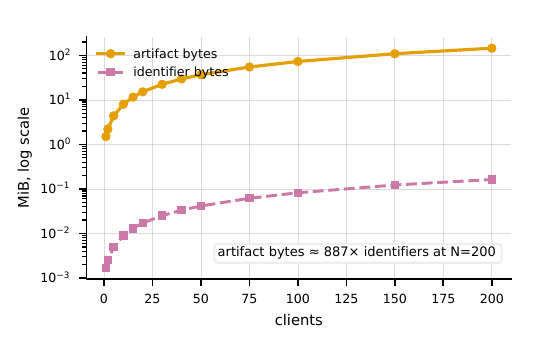}
\caption{Committed identifier bytes and off-chain artifact bytes on MNIST with \(R=12\).}
\label{fig:artifact_vs_identifier}
\end{figure}

Fig.~\ref{fig:artifact_vs_identifier} reports off-chain artifact-volume scaling. At \(N=200\), artifact bytes occupy 144.9 MiB, whereas the committed SHA-256 identifier strings occupy only 0.163 MiB, a ratio of approximately \(887\times\). This supports the design choice of keeping tensors off chain while committing only identifiers and coordination state on chain.

Across all runs, AuditFL matches exactly the corresponding accuracy of Plain FL. 
This is because both protocols execute the same learning rule over the same included updates. In both cases,  final test accuracy ranges from 89.65\% at \(N=1\) to 89.21\% at \(N=200\) clients. 
Note that the variation across client counts results from the 
the differences in non-IID data partitioning (this varies depending on the number of clients). 

\subsection{Controlled Failure Stress} 

We run controlled state-machine stress tests on MNIST with \(N=50\), \(R=12\), three seeds, and artificially induced failure rates from 0\% to 50\%. We test three fault modes.
\begin{inparaenum}[(i)]
\item In \emph{client dropout}, selected clients do not publish an update for the round.
\item In \emph{invalid update}, selected clients publish updates that fail the rule-declared validity checks before aggregation.
\item In \emph{artifact unavailability}, the update identifier is published but the referenced artifact cannot be retrieved during aggregation.
\end{inparaenum}

In these experiments, we use the state-machine backend because the tests concern inclusion, update validity, artifact availability, and round completion rather than EVM execution cost. Artifact-unavailability injections deliberately relax the nominal reliable-storage assumption in order to test whether the clearing layer continues to make progress when referenced updates cannot be retrieved. Very high failure rates, e.g., above 30\%, are included only as extreme stress conditions, not as expected operating points.

\begin{figure}[!t]
\centering
\includegraphics[width=0.95\columnwidth]{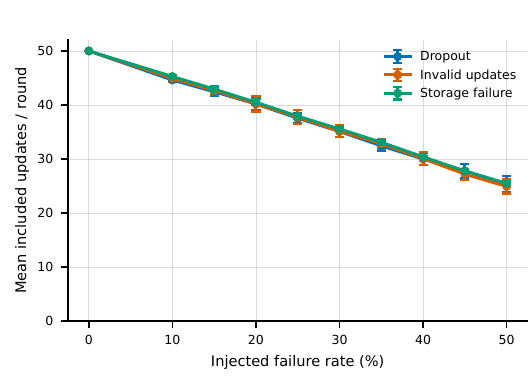}
\caption{Mean included updates per round under injected failures.}
\label{fig:updates}
\end{figure}

\begin{figure}[!t]
\centering
\includegraphics[width=0.95\columnwidth]{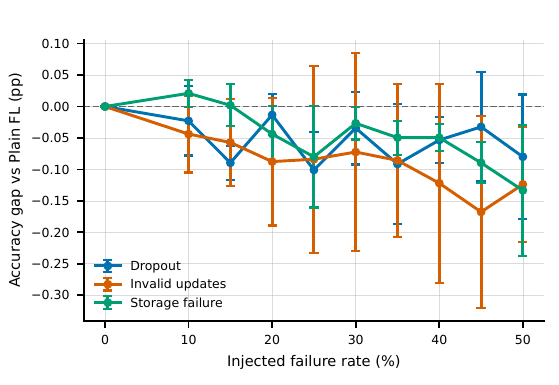}
\caption{Final test-accuracy difference relative to Plain FL under different injected failure modes. The plotted value is AuditFL under failure minus Plain FL, in accuracy percentage points.}
\label{fig:learning}
\end{figure}

Fig.~\ref{fig:updates} shows that client dropout, invalid updates, and artifact unavailability reduce inclusion approximately linearly but do not prevent round finalization. The similar inclusion curves follow from the common aggregation-level consequence of these faults: dropout prevents an update from being published, invalid updates are filtered before aggregation, and unavailable artifacts cannot be retrieved; in each case, the affected contribution is excluded. Even at \(50\%\) injected failures, approximately 25 updates per round remain included.

Fig.~\ref{fig:learning} reports the final test-accuracy difference between each failure run and the corresponding Plain FL run. Negative values mean that the failure run achieved lower final accuracy than Plain FL, while positive values mean that it achieved slightly higher final accuracy. Small positive values can occur because the client data are non-IID, so excluding some updates may occasionally remove noisy or less useful contributions. Overall, the observed changes remain within a few tenths of an accuracy percentage point, even under high injected failure rates. Note that these experiments are intended as stress tests of round completion and learning degradation, rather than 
to compare the fault mechanisms to establish whether one is systematically more harmful than 
the others. 
With only three seeds, we interpret the observed differences as small variations across failure realizations rather than evidence that one fault type is systematically more damaging than another.

\section{Conclusion}

AuditFL is a contract-backed clearing layer for open federated learning. It records ticketed admission, update commitments, round transitions, refunds, rewards, and claims on chain while keeping model artifacts, training, and aggregation off chain. On MNIST with non-IID client partitions, the prototype preserves the Plain FL learning trajectory in clean runs, measured transaction and artifact counts match the analytical model, per-client reserve/publish calls dominate EVM gas, and controlled failures reduce inclusion without preventing round finalization.

The evaluation measures clearing-layer behavior on a local Anvil chain and a local content-addressed artifact store; it does not measure public-network delay or cryptographically enforce aggregation correctness. Future work includes public-network measurements, larger FL workloads, replicated artifact storage, and stronger or verifiable aggregation mechanisms.

\bibliographystyle{IEEEtran}
\bibliography{references_clean}

\end{document}